\documentclass[12pt]{iopart}
\usepackage{iopams}
\usepackage{graphicx}

\newcommand{\bvec}[1]{\ensuremath{\mathbf{#1}}}

\begin{document}

\title[Adatom correlations from first principles]
{What about U on Surfaces? -- Extended Hubbard Models
for Adatom Systems from First Principles}

\author{Philipp Hansmann$^1$, Lo\"ig Vaugier$^1$, Hong Jiang$^2$, and Silke Biermann$^{1,3}$}
\address{$^1$Centre de Physique Th{\'e}orique,
Ecole Polytechnique, CNRS-UMR7644, 91128 Palaiseau, France\\
$^2$Beijing National Laboratory for Molecular Sciences, College 
of Chemistry 
and Molecular Engineering, Peking University, 100871 Beijing, China\\
$^3$Japan Science and Technology Agency, CREST, Kawaguchi
 332-0012, Japan}


\ead{philipp.hansmann@cpht.polytechnique.fr}

\begin{abstract}
Electronic correlations together with dimensional constraints lead to some
of the most fascinating properties known in condensed matter physics. 
As possible candidates where these conditions are realized, semiconductor 
(111) surfaces and adatom systems on surfaces have been under investigation 
for quite some time. However, state-of-the-art theoretical studies on these 
materials that include many body effects beyond the band picture are  
rare. First principles estimates of inter-electronic Coulomb interactions
for the correlated states are missing entirely, and usually these interactions are
treated as adjustable parameters. In the present work, we report on calculations of
the interaction parameters for the group IV surface-adatom systems in the 
$\alpha$-phase series of Si(111):{C, Si, Sn, Pb}. For all systems investigated,
inter-electronic Coulomb interactions are indeed large compared to the 
kinetic energy of the states in question. Moreover, our study reveals
 that intersite interactions cannot be disregarded. 
We explicitly construct an extended Hubbard model
for the series of group IV surface-adatom systems on silicon, which can
be used for further many-body calculations.
\end{abstract}

\maketitle

\section{Introduction: $\alpha$-phases of Si(111)}
Strongly correlated electron systems have become the focus of an ever-growing activity in the last decades. One of the reason for the great popularity of such systems can be found in the highly bundled variety of fascinating physics, which manifests itself in rich phase
diagrams and highly non-trivial ground states. Specifically, systems which inherit a dimensional constraint on top of strong electronic correlations, display exotic ground states such as high T$_{\rm C}$ superconductivity, fractional quantum Hall effects or unusual magnetic states.\cite{imada98} 
Moreover, the technological progress allows for controlling growth processes and, hence, the atomic structure of thin films and interfaces on the nanometer scale opening the path for actual materials design. While the experimental techniques developed, theoretical methods also experienced a remarkable evolution within the past two decades aimed at an \emph{ab initio} (i.e. a parameter-free) treatment of quantum many body systems.

A natural question arises about how the strength of correlations can be estimated. The common wisdom is that the family of transition metal oxides (specifically those from the 3d series) and rare earth compounds should be considered as the archetypical example of a correlated electron system. The reasoning is that electrons in partially filled d-shells are, due to a non optimal screening of the core charge, dragged towards the nucleus and, hence, highly localized. Consequently, they ``interact more'' with one another and the energy scale of local atomic-like physics becomes comparable to that of the hopping/delocalization. This last conclusion is, however, a more general statement and poses no restrictions upon the quantum numbers of the states in consideration: it merely relies on the comparison of the kinetic and the potential Coulomb energy. 

It has been suggested some time ago, in 1974, by Tosatti and Anderson \cite{tosatti74} that by this argument, one should be able to observe low dimensional correlated physics on semiconductor, silicon or germanium (111), surfaces. The states for which the correlation criterion above would apply arise from the dangling bonds (i.e. unsaturated $sp^3$-bonds) at these specific surfaces.
Since this proposal, such surfaces have been explored experimentally (see e.g. \cite{uhrberg85,grehk93,weitering93,carpinelli96,carpinelli97,weitering97,lay01,pignedoli04,upton05,modesti07,cardenas09,zhang10,tournier11}) and from the theoretical side (many of them by means of density functional theory - see e.g. \cite{kaxiras90,brommer92,santoro99,hellberg99,aizawa99,profeta00,shi02,shi04,profeta05,profeta07,schuwalow10,chaput11,li11}). The first impression from the results are discouraging with respect to the correlation scenario: more often than not, one finds simply a structural rearrangement of the surface in order to saturate the dangling bonds to an energetically more favorable configuration. This trend results in a wide variety of -- sometimes large -- surface unit cells: The pure silicon (111) surface, e.g., reorders to an (almost) band insulating state with a lattice constant of $\approx 22$\AA. Luckily it has turned out, however, that certain adatom systems (i.e. surface plus adsorbed atoms) do not suffer from such structural rearrangement but remain undistorted. Strangely enough some of these systems are adatom systems with group IV adsorbed atoms (i.e. isoelectronic to silicon). The structures we will report on in this manuscript belong to this so called $\alpha$-phase family and are members of the series Si(111):$\{{\rm C, Sn, Pb}\}$. The $\sqrt{3}\times\sqrt{3}$ surface unit cell contains one adatom and three silicon/germanium atoms in the top layer so that the coverage can be interpreted as 1/3 monolayer (ML). As a side remark we note that the actual coverage -- and experimental control of it -- is a crucial issue. The physics dramatically depends on the electron count associated with a specific coverage \cite{hellberg99,chaput11}. 
At 1/3 coverage, the adatom occupies the T$_4$-position and effectively saturates three dangling bonds of the topmost silicon layer albeit contributing an additional unbound dangling bond electron itself. In the following we will 
present our considerations 
of the Si(111):$\{{\rm C, Sn, Pb}\}$ series, along with
the hypothetical structure Si(111):Si which does not exist but will still be helpful for identifying trends within the series. 

Experimentally the $\alpha$-phase structures are not yet comprehensively explored and in some cases reports have been even controversial\cite{hellberg99,chaput11}. It is, however, unquestionable that these materials show a remarkable variety of interesting physics including comensurate charge density wave (CDW) states \cite{carpinelli96,carpinelli97,lay01} and even isostructural metal to insulator transitions (MIT)\cite{modesti07}. While there have been also many studies and model calculations on the theoretical side, systematic studies are rare (\cite{santoro99}) not even to mention a unified understanding. The work we present here is intended to contribute to such understanding from the perspective of state-of-the-art \emph{ab initio} calculations for electronic correlation effects. More specifically, we will report calculated values for the parameters of the Coulomb interaction for the series  Si(111):$\{{\rm C, Sn, Pb}\}$. Of these compounds specifically the Si(111):Sn system has been already subject to electronic structure studies beyond density functional theory (DFT) within the local density approximation (\cite{schuwalow10,li11}) where the interaction parameters remained, however, adjustable and set by hand. From our results, it will become furthermore clear that when folding down/projecting the problem to the single band case, intersite interactions with the six nearest neighbor sites should not be disregarded as they turn out to be large compared to the onsite interaction (and just by the trigonal structure with six nearest neighbors they will be even more important than in the cubic case). 
The importance to go beyond the standard Hubbard model by means of longer range correlation effects was, in fact mentioned and partially studied in previous works \cite{schuwalow10, li11} and intersite interaction terms have been considered as parameters in Hartree Fock calculations\cite{santoro99}. Our results, confirming such importance, and subsequent analysis of the material-specific extended models can be expected to significantly contribute to the understanding e.g. of the charge ordered phases of Ge(111):Sn \cite{carpinelli97} and Ge(111):Pb \cite{carpinelli96}, where just a simple nesting scenario is considered not to be sufficient \cite{santoro99} to explain the experimental observation of a CDW ground state. We mention in passing that the charge ordered phase of Ge(111):Sn has been considered by Schuwalow et al. \cite{schuwalow10}. The inequality of the Sn sites in the charge ordered phase was imposed already as structural input. Complementary to this study we suggest to investigate the CDW instability from a generic starting point and check how correlation effects, and instabilities in the electronic structure, possibly even conspire with structural responses (an issue known also for correlated oxides, e.g. in the case of V$_2$O$_3$).
Furthermore, our results may even help to predict possible ordered phases for the experimentally rather unexplored Si(111):C compound, as it was first attempted by Profeta and Tosatti \cite{profeta05}.

\section{Methods: the constrained Random Phase
Approximation (cRPA)}

In order to perform \emph{ab initio} calculations for strongly correlated electron systems the formulation, and subsequent solution, of a lattice model has turned out to be a successful strategy. In this ``tight binding'' philosophy the energy bands are understood in terms of atomic-like quantum numbers and hopping processes on the lattice. To construct a lattice Hamiltonian we write down a single particle operator describing such hopping, i.e., kinetic energy of the electrons and, as a separate term, the two-particle operator of inter-electronic Coulomb interactions. In practice, this is usually done by construction of a multi-band Hubbard-like Hamiltonian which simplifies the full Coulomb interaction into an energetic cost of doubly occupied states or -- in the multiorbital case -- multiply occupied sites. This additional cost corresponds to the difference between the electron affinity and the ionization energy and is usually parametrized by the famous Hubbard $U$ (which, in the multi-band case, acquires a matrix form including also Hund's coupling as well as higher multipole orders of the Coulomb interaction). The multi-orbital Hubbard model reads:\\
\begin{equation}
\hat{H}=\sum_{iljm\sigma}t_{il,jm}\hat{c}^\dagger_{il\sigma} \hat{c}_{jm\sigma}+\sum_{ilmno\sigma\sigma'}U_{lmno}\hat{c}^\dagger_{il\sigma} \hat{c}^\dagger_{im\sigma'} \hat{c}_{in\sigma'} \hat{c}_{io\sigma}
\label{hubbard}
\end{equation}
where $\hat{c}^\dagger_{il\sigma}$($ \hat{c}_{il\sigma}$) creates (annihilates) an electron with spin $\sigma$ and orbital index $l$ at lattice site $i$; $t_{il,jm}$ is a hopping amplitude between lattice sites $i$ and $j$ and (Wannier--)orbitals $l$ and $m$; finally, $U_{lmno}$ denotes the full parametrization of the local Coulomb interaction. The first, kinetic/hopping- part of the Hamiltonian is usually derived from the DFT Kohn-Sham eigenstates and bandstructure while, in a subsequent step, the full interacting Hamiltonian is solved. However, the full basis set of a standard DFT calculation involving up to hundreds of wave functions is too large for a many-body treatment (and irrelevant for the low energy physics), motivating the derivation of \emph{effective low energy Hamiltonians}. Such a construction involves downfolding\cite{lowdin51,andersen00, aichhorn09} onto states in a certain energy window around the Fermi energy. 

At a second glance, an obvious question arises: while the kinetic part of the Hamiltonian is usually derived in a straightforward manner from the full DFT result, there remains the task to handle the interaction part of the problem. In the DFT framework, the inter-electronic Coulomb potential $\propto 1/|r-r'|$ is replaced in a mean field like spirit (translated to an effective single particle potential) in order to solve only single particle Schr\"odinger equations. 
In order not to leave the interacting part of the Hamiltonian as an adjustable parameter, there have been several attempts to calculate these quantities from first principles. Up to now most of the procedures rely on constrained density functional approaches~\cite{dederichs84}. 
Such ``constrained LDA'' approaches~\cite{mcmahan88,hybertsen89, gunnarsson89,gunnarsson90,anisimov91} are based on the observation that the energy of a system with increased or reduced particle number is in principle accessible within density functional theory. $U^{\rm (cLDA)}$ is then defined as the second order derivative of the total energy 
with respect to the local occupation number of the correlated states.

In 2004, however, an approach derived in the same Wilson-like renormalization philosophy as the downfolding/projection for the kinetic energy part of the Hamiltonian has been proposed by Aryasetiawan and co-workers\cite{aryasetiawan04}: the so called constrained random phase approximation (cRPA).
Its central idea is to construct a \emph{partially} screened Coulomb interaction, that has the following property: when used as a ``bare'' interaction in conjunction with the low-energy single-particle part, screening {\it within this low-energy space} leads to a fully screened Coulomb interaction that is equal to the one of the initial system. Matrix elements of this effective partially screened interaction can thus be interpreted as the bare interaction within the low-energy description, that is the Hubbard $U$ and Hund's rule $J_{\rm H}$ needed for the effective Hamiltonian.\\

\begin{figure}%
\centering
\includegraphics[width=\textwidth]{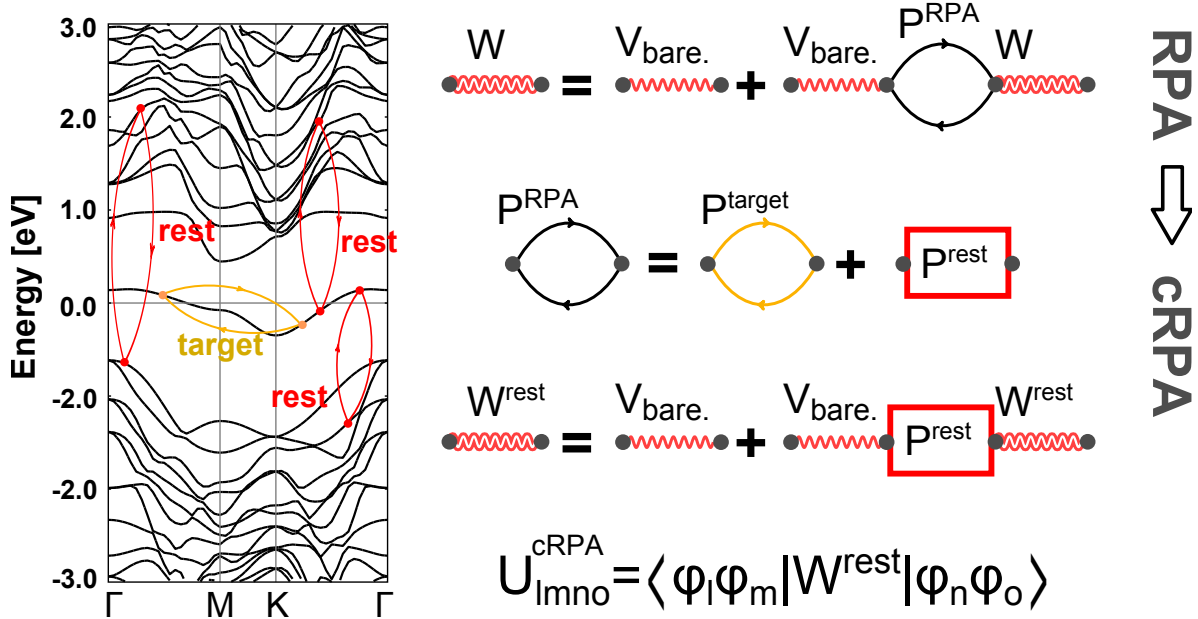}%
\caption{(Color online) Sketch of particle-hole excitations contributing to the polarization operator screening the bare Coulomb interaction. In the constrained RPA procedure we exclude intra-surfacestate particle-hole excitations (yellow/light gray). More specifically, we ommit any convolution of intra-surfaceband particle or hole propagators.}%
\label{bubbles}%
\end{figure}

For the case of the silicon (111) surfaces we can nicely exemplify the cRPA procedure. In Fig. ~\ref{bubbles} we show on the left hand side the bandstructure for the Si(111):Sn representative for all $\alpha$-phase systems. In the RPA scheme, the screened interaction is calculated by approximating the polarization within the simplest Feynman diagram for the polarization in the Dyson equation for the interaction: the bare bubble. 

The philosophy of the constrained RPA is to exclude certain particle-hole (ph) excitations, in order to avoid a double counting in subsequent treatment of the effective model. The exclusion condition can be tailored to the specific need and nature of the subsequent treatment: An energy window could be selected for which we exclude internal particle-hole diagrams, a certain band index can be choosen to exclude intraband particle-hole diagrams or, more sophisticated one can pose a condition that particle-hole excitations of localized states should be excluded. In our present case, due to the disentangled, simple surface state all these conditions are equivalent: We exclude intra-surface state particle-hole excitations.

On the right hand side of Fig.~\ref{bubbles}, we sketch the idea of cRPA which consists in separating the polarization operator into a part which acts inside the energy window and the rest (as it is also indicated by the particle-hole excitations in the band structure plot). 
\begin{eqnarray} 
  P&=& P^{\rm target}+P^{\rm rest},
\label{polsep}
\end{eqnarray}
Here, 
$P^{\rm target}$
denotes the polarization involving only processes \emph{within} the single \emph{target} band around the Fermi energy. Hence, $P^{\rm rest}=P-P^{\rm target}$ is a constrained polarization, in which the screening contributions of the target space have been projected out.
Such constrained polarization leads to the partial dielectric function $\varepsilon^{\rm rest}$ 
\cite{GW_Ferdi,hedin65}. 
\begin{eqnarray} \label{dielectric}
  \varepsilon^{\rm rest}(1,2) &=& \delta(1-2) - \int d3 \, P^{\rm rest}(1,3)v(3,2).
\end{eqnarray}
(where the numbers represent space and frequency coordinates in a shorthand notation).
The \emph{partially} screened interaction $W^{\rm rest}$ can then be defined as follows: 
\begin{eqnarray} \label{wrest}
  W^{\rm rest}(1,2) &\equiv& \int d3 \, \varepsilon_{\rm rest}^{-1}(1,3) v(3,2).
\end{eqnarray}
Since screening of $W^{\rm rest}$ with processes from within the target space recovers the fully screened interaction $W$, it is justified to interpret the matrix elements of $W^{\rm rest}$ in the localized Wannier basis as the interaction matrices. It should be stressed that in equation (\ref{wrest}) $W(\omega)$, and therefore $U(\omega)$, is still a function of frequency/energy! 
The effects of such dynamically screened interactions are currently
receiving much attention \cite{werner-natphys,casula12}.
In this report we will restrict ourselves, however, to report on the static values $U(\omega=0)$
\begin{eqnarray}
U_{lmno}  = \langle \phi_{l} \phi_{m} | W^{\rm rest}(0) | \phi_{n} \phi_{o}\rangle \label{umatrix}
\end{eqnarray}

We employ the cRPA method as recently implemented within the linearized augmented plane wave framework in Ref.~\cite{vaugierPHD,vaugier12}. The constrained polarization $P^{\rm rest}$ is deduced from the Kohn-Sham eigenstates of the one-particle Hamiltonian, i.e., obtained within DFT in the electronic structure code Wien2k~\cite{wien2k}. Wannier functions are constructed from projected atomic orbitals promoted to the status of true Wannier functions through an orthonormalisation procedure~\cite{aichhorn09}.

Before we turn to the results, we emphasize that the polarization, and thus the dielectric tensor (\ref{dielectric}) $\varepsilon(\omega,\bvec{r},\bvec{r}')$ (where $\omega$ is energy and $\bvec{r},\bvec{r}'$ are position vectors of the interacting charges) can be, depending on the specific system, an involved object. The simplification of this object to a dielectric function, neglecting energy-dependence and proposing translational symmetry of a continuum, $\varepsilon(|\bvec{r}-\bvec{r}'|)=\varepsilon(\bvec{q})$ or even a dielectric constant $\varepsilon$ may be a dramatic oversimplification of the real quantity. It has been pointed out by van den Brink and Sawatzky \cite{vdbrink00} that, specifically for dimensionally constrained systems, screening processes can involve highly non-trivial physics when considering lenght scales comparable to inter-atomic distances. In cRPA these so called \emph{local field} effects are, on the level of the random phase approximation, included and we will later compare our results also to a static continuum approximation.

\section{Results \& Conclusions}
\begin{figure}%
\includegraphics[width=\textwidth]{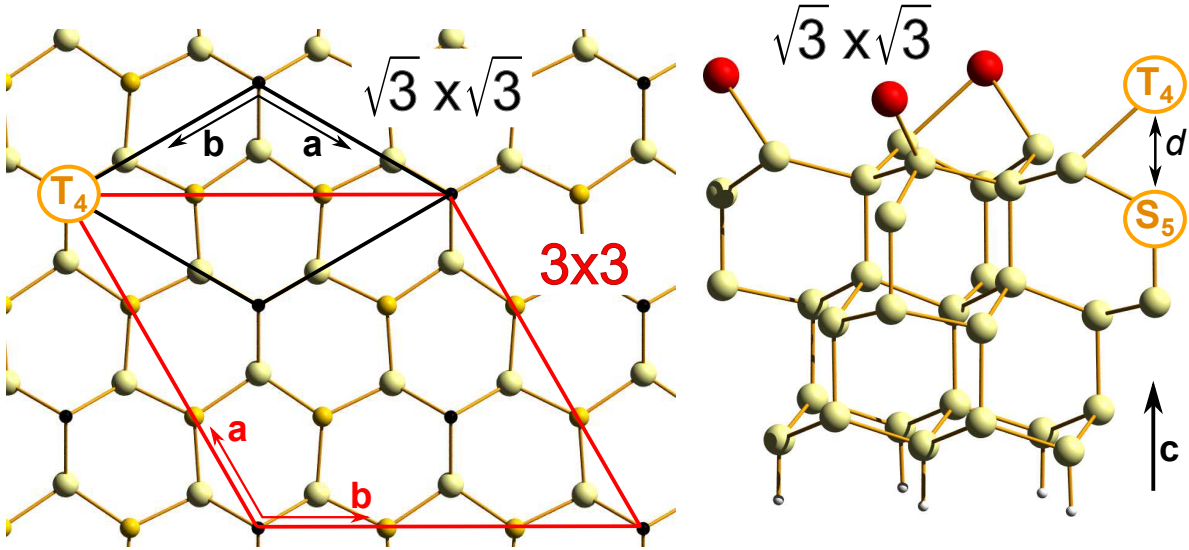}%
\caption{(Color online) Left hand side: Top view on the Si/Ge(111) surface and unit cell dimensions for the $\alpha$- $\sqrt{3}\times\sqrt{3}$ phases and the $3\times 3$ phase. Different Si/Ge layers are indicated by size/color. Right hand side: Side view of the unit cell of the $\alpha$- $\sqrt{3}\times\sqrt{3}$ phase as it was used for the calculations indicating the important adatom sites $T_4$ and $S_5$ (adatoms in red/dark gray)} %
\label{unitcell}%
\end{figure}

Starting from the bandstructure results we will discuss in the following section the electronic structure of the $\alpha$-$\sqrt{3}\times\sqrt{3}$ phase 1/3 ML coverage adatom systems for the Si(111):$\{{\rm C, Si, Sn, Pb}\}$ series.

Studying surfaces within standard density functional codes calls for a structural input by means of a ``slab'' geometry which we show in Fig.\ref{unitcell} on the right hand side. We decided to employ the same type of unit cell as the authors of Ref.\cite{schuwalow10} used for their Si/Ge(111):Sn study. It consists of three bilayers of silicon which are saturated by hydrogen atoms on the bottom side. On the top side the adatoms occupy the T$_4$ position. Exceptional in this sense is the C adatom system, in which the carbon atom actually occupies 
a subsurface position, 
the S$_5$ position, while T$_4$ is occupied by a silicon atom. Resulting from this geometry each adatom occupying a T$_4$ site ''saturates'' the DBs of three underlying silicon atoms (see Fig.\ref{unitcell} left hand side) while, at the same time, naturally adding one DB by themselves (note that, in contrast, group III adatoms simply saturate the DBs leaving the surface band-insulating). These slabs are then separated by a sufficiently large vacuum region (of the order of $10$\AA) in the c-direction.

The 
calculations have been performed in three subsequent steps: i) structural relaxation within a projector-augmented-wave (PAW) basis set as implemented in the Vienna ab initio simulation package (VASP) code \cite{vasp}, ii) the calculation of the electronic structure and subsequent processing within the full potential linearized augmented plane wave (LAPW) Wien2k code \cite{wien2k},
iii) the cRPA calculation within the implementation of Ref. (\cite{vaugierPHD,vaugier12}).
Convergence with respect to the size of the vacuum region as well as density of the lateral k-point mesh was checked. In this way we reproduce in step i) structural input as it was obtained by previous studies of Si(111):C (see \cite{pignedoli04}) and Si(111):Sn (see e.g. \cite{schuwalow10}). 

\begin{figure}%
\includegraphics[width=\textwidth]{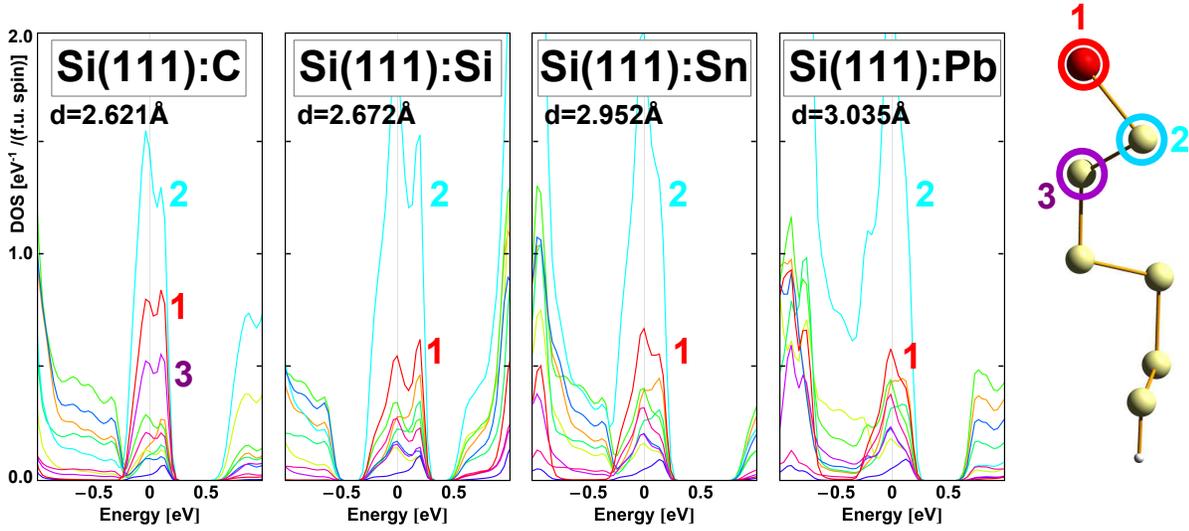}%
\caption{(Color online) Atom-resolved densities
of states (DOS) within $2$eV of the Fermi energy (which is set to $0$) calculated with DFT LDA. 
Each DOS has been weighted with the multiplicity of the atom in the unit cell. The main contributions can be traced back to the upper two layers (positions 1 and 2) of the surface. Specifically in Si(111):C also the sublayer position 3 contributes.}%
\label{DOSgrid}%
\end{figure}

\paragraph{Single Particle DOS \& Bandstructure}
We start our discussion of the electronic structure results by reporting on the single particle density of states (DOS) shown in Fig.\ref{DOSgrid}. In the figure we show the DOS in a narrow energy window of $\pm1$ eV around the Fermi energy ($\varepsilon_{\rm F}=0$ eV). Immediately evident is a narrow peak of roughly $\approx0.5$ eV width which has appeared  inside the semiconducting gap of bulk silicon. This feature has to be attributed to electronic states absent in the bulk and arise from a half-filled narrow surface-band. 

At a closer look at the DOS, however, we realize that the term ``surface-band'' must be used carefully: The plots in Fig.\ref{DOSgrid} are resolved as partial density of states (pDOS) with respect to the contribution of the different atoms in the unit cell weighted by the multiplicity factor of the respective atom. 
Before further discussion, a word of caution is in order here:
the quantitative composition of the pDOS depends of course on the details and choice of the DFT projectors. 

More specifically, in Wien2k,  the pDOS is calculated by projecting Kohn-Sham wave functions into the LAPW basis in the muffin-tin region so that the pDOS results depend on the choice of the muffin-tin radius. Nonetheless we can, at least on a qualitative level, take a message from the pDOS plots: Not only the adatom contributes to the new ``surface-state'' which is, in reality rooting deeper inside the substrate than what may have been anticipated. It should probably best be thought of as a molecular orbital (MO), i.e. a linear combination of orbitals of the adatom and substrate atoms. 
In fact, the contribution of the adatom and the three topmost silicon atoms of which it saturates the dangling bonds contribute approximately equivalently (ratio of turquoise to red is roughly 3:1) for most of the MO. The center of gravity of such MO will, hence, be shifted away from the adatom site down into the substrate due to the lack of z-inversion symmetry, i.e., overlap into the vacuum. Let us in this context also mention, without going to details, that the composition of the MO depends crucially on the structure as for example the distance between the adatom and the underlying substrate (values for this T$_4$ to S$_5$ distance are reported in Fig.~\ref{DOSgrid}). In line with the MO language, specifically the nature of highest occupied (HOMO) or lowest unoccupied (LUMO) state depends on the structure and we remark that the Si(111):C system seems to be qualitatively different from the rest of the series. It will, indeed, turn out to be the compound with the most localized surface-states. 

This qualitative argument of a MO interpretation of the surface-state is also reflected in the Wannier function construction (compare Fig.5 of Ref. \cite{schuwalow10}) and, hence, in the evaluation of the interaction parameters. On the other hand, while this \emph{apical} extension seems larger than one might have expected, the lateral (planar) extension of the orbital should not be too large compared to the nearest neighbor distances on the trigonal adatom lattice as the kinetic energy (or hopping) is only of the order of $0.05$eV as we will see - figuratively speaking the orbital resembles a ''carrot'' sticking in (or growing out of) the surface.

\begin{figure}%
\includegraphics[width=\textwidth]{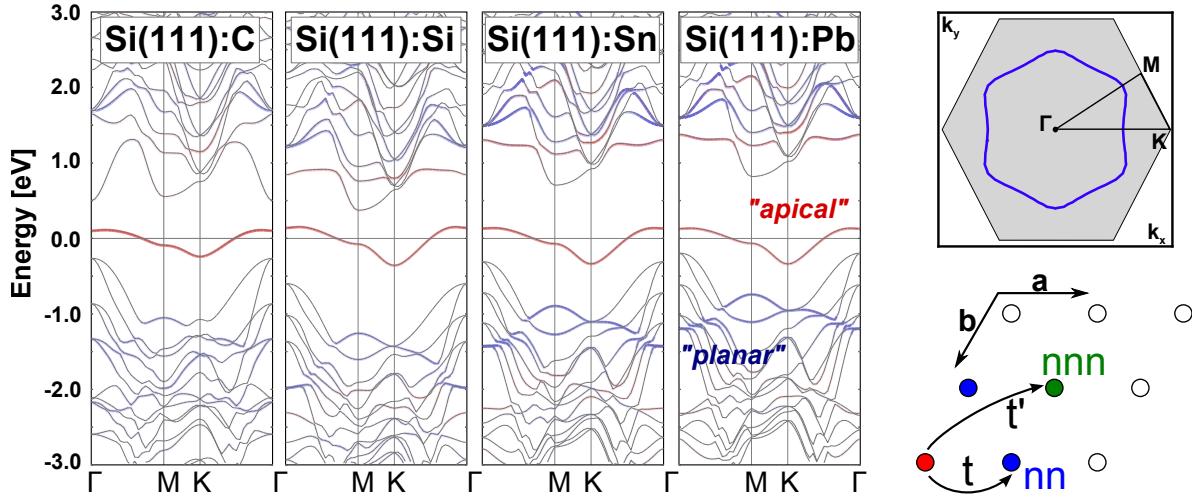}%
\caption{(Color online) Left hand side: Bandstructures of the $\alpha$- $\sqrt{3}\times\sqrt{3}$ phases for group IV adatoms\cite{silicon}. The color of the bands denotes their respective orbital character. Red color indicates a $p_z$-like ``apical'' character, while the blue color denotes $p_{x,y}$-like character. Right hand side: Sketch of Brillouin zone and Fermi surface originating from the surface band (top) and sketch of nearest neighbor and next nearest neighbor configuration on the trigonal adatom lattice.}%
\label{fatbands}%
\end{figure}

Let us now turn to the bandstructure results for the $\alpha$-phase series, reported in Fig.\ref{fatbands} in the left panels. On the right hand side of the figure we show, in the upper part, a sketch of the Brillouin zone (in reciprocal Cartesian coordinates $k_x$ and $k_y$) including important high symmetry points and the qualitative form of the 2D Fermi surface. In the lower part we show a sketch of the trigonal lattice formed by the adatom sites, indicating also the geometries of nearest and next nearest neighbor coordinations.

The most striking observation from the band structure plots in Fig.\ref{fatbands} is, that for all considered systems the surface-state in the semiconducting gap (as previously discussed for the DOS) is indeed responsible for a \emph{well separated single band around the Fermi energy}. Turning back to the plots of the DOS we can now clearly see that the closing of the small gap (we can now see that it is an indirect gap) below the Fermi energy from Si(111):Si to the heavier adatoms not indicates a qualitative difference by means of separation from the bulk states. The surface states around the Fermi energy are not entangled with underlying bulk states in \emph{any} of the adatom systems. In the plots we have encoded additional (qualitative) information by means of the coloring of the bands: In red (gray) we plot the contributions stemming from the $p_z$-orbital of the adatom while we plot the adatom $p_{x,y}$-character in blue (dark gray). What we can learn from these colored plots is that, even though the actual MO composition might be complicated, the half filled surface band has a clear cut ``apical'' (i.e. carrot like) character. The ``planar'' $p_{x,y}$ states, on the other hand, are responsible for forming the covalent bonds to the substrate silicon and are, hence, either completely filled or completely empty.  

The simple structure of the half filled band suggests to write down the simplest possible tight binding dispersion relation which we formulate up to next-nearest neighbor hopping (as indicated in Fig.\ref{fatbands} on the lower right hand side):

\begin{eqnarray*}
\fl \varepsilon_{\bvec{k}}=2t\cdot\left(\cos(k_x)+2\cos(k_x/2)\cos(\sqrt{3}/2 k_y)\right)\\
+2t'\cdot\left(\cos(\sqrt{3}k_y)+2\cos(3k_x/2)\cos(\sqrt{3}/2 k_y)\right)\\
\label{TBHam}
\end{eqnarray*}

where $t$ and $t'$ are the amplitudes for nearest and next-nearest neighbor hopping processes.
It turns out, that this real space cut still allows for qualitatively reasonable fitting and  yields for Si(111):$\{{\rm Si, Sn, Pb}\}$ values of $t\approx 0.05$ eV and $t'\approx 0.4t$ eV. As already inferred from the pDOS plots the peculiarity of the Si(111):C system is also reflected in the tight binding analysis and yields a smaller value for hopping $t$ and, more importantly also a smaller $t'/t$-ratio of $\approx 0.3$ which again points towards a more localized MO state. Overall, the small hopping amplitudes and resulting bandwidth actually fit our context of an \emph{apical} orbital nicely. In this case the hopping/hybridization is of $\pi$-character which yields a significantly smaller hopping amplitude as opposed to a $\sigma$-type hybridization.

\begin{table}
\caption{\label{uvaluestab} Values of the bare (V) and screened (U) values for on- and intersite nearest neighbor (nn) interaction parameters (Compare Fig.\ref{uvalues}).}
\footnotesize\rm
\begin{tabular*}{\textwidth}{@{}l*{15}{@{\extracolsep{0pt plus12pt}}l}}
\br
[eV] &Si(111):C&Si(111):Si&Si(111):Sn&Si(111):Pb\\
\mr
$V$&$6.0$&$4.7$&$4.4$&$4.3$\\
$U$&$1.4$&$1.1$&$1.0$&$0.9$\\
$V^{\rm nn}$&$2.8$&$2.8$&$2.7$&$2.8$\\
$U^{\rm nn}$&$0.5$&$0.5$&$0.5$&$0.5$\\
\br
\end{tabular*}
\end{table}

\begin{figure}%
\includegraphics[width=\textwidth]{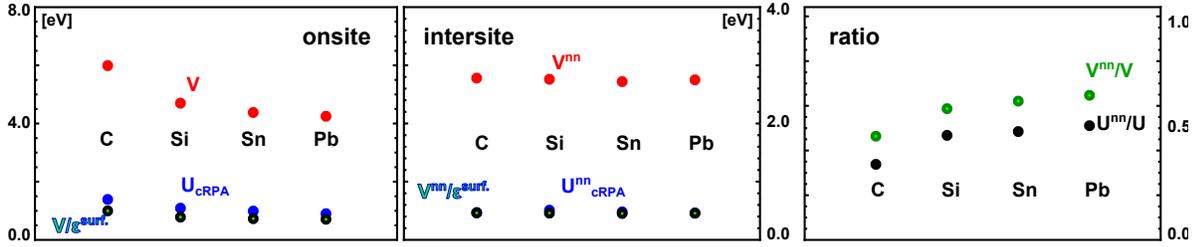}%
\caption{(Color online) Results of the constrained RPA calculation for the $\alpha$- $\sqrt{3}\times\sqrt{3}$ phases for group IV adatoms\cite{silicon}. Left panel: Comparison of bare \emph{onsite} Coulomb interaction (red/light gray) 
and the screened value U given by i) cRPA (blue/dark gray) and ii) continuum limit estimate (green/open circle); Center panel: Same comparison for nearest neighbor \emph{intersite} interaction; Right panel: Comparison of the ratio between intersite and onsite interactions for the bare ($V^{\rm nn}/V$) and screened cRPA ($U^{\rm nn}/U$) values (Compare Tab.\ref{uvaluestab}).}%
\label{uvalues}%
\end{figure}

\paragraph{Interaction parameters}
Let us, finally, turn to the results for the cRPA calculations. For the projection onto the low energy single band Hamiltonian we have chosen to project in a suitable energy-window framing the surface band around the Fermi energy onto the $p_z$-character of the adatom. By construction, the resulting Wannier function is carrying implicit information about all states (not only the $p_z$) which contribute to the surface-band. This implicit information is reflected in the tails of the resulting Wannier function (as it can be seen e.g. in Fig.5 of Ref.\cite{schuwalow10}) and it actually does not depend on which orbital one projects in the first place as long as it has a finite contribution to the band of interest. Then, in a first step, the bare interaction parameters are calculated by means of explicit evaluation of the radial integrals over the Wannier function. In the second, actual cRPA step, the dielectric tensor is evaluated as described in the previous section and the screened quantities - onsite and intersite interactions - are calculated. A summary of these results can be found in Tab.\ref{uvaluestab} and in Fig.\ref{uvalues}. 

The bare-onsite interaction parameters are between $6.0$ eV for Si(111):C and $4.3$ eV for Si(111):Pb decreasing monotonously within the series. 
These values are small compared to the values of atomic orbitals, consistently with our picture of the surface state being of MO
nature rather than an atomic adatom orbital. The cRPA screened values for the series are smaller roughly by a factor $4-5$ and range from $1.4$ eV for Si(111):C to $0.9$ eV for Si(111):Pb, again monotonously decreasing. Comparing these values to the bandwidth of $\lesssim 0.5$ eV makes it evident that electron-electron interaction yields potential energies on the same scale as the kinetic hopping energy. Remembering the critical U values of the Hubbard model, one might interpret the cRPA result as a clear evidence for the Mott physics of the system. This is, however, a premature conclusion for a reason that can be seen already from the bare interaction values: the intersite contributions. 
Shown in the last two lines of Tab.\ref{uvaluestab} we report on the bare and screened nearest neighbor (nn) interaction parameters. In Fig.\ref{uvalues} middle and right panel we visualize the intersite values and the intersite/onsite ratio respectively. Two observations have to be emphasized:  i) 
The bare nn-intersite interactions are half the size compared to the onsite interaction parameters and ii) the values of the bare and screened intersite interaction parameters barely change throughout the materials series.

Apparently the differences of the Wannier function in the different materials affects the intersite- less than the onsite overlap integrals. The reason for this is that for all the series the intersite overlap is so small that the Coulomb energy almost corresponds to the electrostatic energy of two point charges. We can quantify this argument further 
as follows:
The expectation value $\langle \frac{e^2}{r_{\rm rel}} \rangle$
of the Coulomb interaction of the two charges, expressed in terms
of their relative coordinate ${\bf r}_{\rm rel}$, can be related
to a rescaled hydrogen atom. Indeed, up to the sign (repulsion
instead of attraction) this problem ressembles a hydrogen atom
with effective Bohr radius 6 $\AA$.
Relating the potential energy to the total energy via the virial
theorem, $\langle E^{\rm tot.}\rangle=1/2 \langle V\rangle$, we
obtain:
\begin{eqnarray}
\langle \frac{e^2}{r_{\rm rel}} \rangle = \frac{1}{12}
|V_{\rm pot}^{\rm H atom}| = \frac{1}{12} 2 |E_{\rm ground state}^{\rm H atom}|=2.3 eV
\end{eqnarray}
which roughly matches the value of our bare intersite interaction parameters. 

The corresponding inter- vs. onsite interaction ratio from Si(111):C to Si(111):Pb hence shows an increasing trend in the series. 
Ratios of the order of $U^{\rm nn}/U\approx0.5$ suggest that the system is very much influenced, if not dominated, by intersite- rather than onsite electronic interaction. As we have mentioned in the introduction part, these results suggest to revisit specifically the charge ordered phases of Ge(111):Sn, Ge(111):Pb, as well as predictions for possible ground states in Si(111):C. More specifically, we will proceed in future work to investigate the extended Hubbard model taking into account also intersite nearest neighbor interaction terms with the goal to come closer to a unified picture of the variety of ground states of the different $\alpha$-phase compounds.

In addition to the cRPA values we also show (left and middle panel of Fig.\ref{uvalues}) the values estimated from the bare interaction and the continuum limit approximation for a dielectric constant as it was mentioned in the previous section. The value of the continuum approximation for the screening of charges on a surface can be found by an image-charge estimate to be 
half of the bulk value $\varepsilon_{\rm surf.}=\frac{\varepsilon+1}{2}$. We observe that the values obtained within the continuum approximation are in quite good agreement and consistent with the cRPA calculated values for the onsite as well as for the intersite interaction values. The validity of the continuum approximation, specifically for the intersite interactions can be easily understood: comparing the inter-atomic distances within the silicon bulk substrate ($\lesssim 2.5$\AA) to the distances of the adatoms ($\approx 6$\AA) we infer that the latter are sufficiently \cite{vdbrink00} large in order not to suffer from local field corrections.
From a technical point of view, this agreement is an additional {\it a posteriori} confirmation
that the calculated values should be converged with respect to the k-point resolution and energy cutoff of the particle-hole bubbles.\\

Finally, we should remark that in the present calculations possible effects of spin orbit coupling (SOC) have not been taken into account explicitely. In fact, such effects are not expected to severely change the calculation of interaction parameters. A local SOC would mix in states belonging to the same subspace of total angular momentum J to the single orbital of our target basis. While this could certainly affect the expectation value of observables reflecting local moments like, e.g., magnetization, it is not expected that such effect would change the spread and hence the Coulomb radial integrals of the projected orbital. The situation would be different if we had more than one degenerate bands around the Fermi energy - then even small SOC could lead to delicate splittings with big effects\cite{martins12}. Follow-up  studies should therefore carefully consider wether or not SOC, specifically for the heavier adatoms like Sn and Pb, could play an important role.

\paragraph{Conclusion}
In conclusion we have calculated \emph{ab initio} values for the inter-electronic interaction parameters for the series of $\alpha$-phase group IV adatom system on the silicon (111) surface. Our results show, that for the entire series the order of magnitude of the onsite Hubbard-interaction parameter U is of comparable and even larger size than the kinetic energy of the surface-adatom states. Moreover, our results have clearly shown that intersite interaction (specifically nearest neighbor interaction) should not be neglected and that many-body methods should consider an extended Hubbard type Hamiltonian. We argue that, specifically in the trigonal lattice symmetry, the intersite terms might be dominating the ground state physics of the systems.
Instabilities of the electronic structure towards a charge ordered phase (as it is seen in Ge(111):Pb or Ge(111):Sn) will sensitively depend on non-local interaction terms. As an outlook we point out that a possible instability in the electronic structure can also back couple to the lattice structure. This aspect might be especially interesting in the case of Ge(111):Sn where the $3\times3$ charge ordered phase is also associated to structural rearrangement (by means of a vertical distortion of the Sn-adatom).  

\section{Acknowledgements}
We acknowledge useful discussions with the authors of
(\cite{tournier11}), as well as with
F.~Aryasetiawan, T.~Ayral, M.~Imada, T.~Miyake, K.~Nakamura, and G.~Sawatzky.
This work was supported by the French ANR under project SURMOTT
and IDRIS/GENCI under project 129313.


\begin{thebibliography}{10}
\bibitem{silicon} The structure with silicon as an adatom does not exist. It is calculated/presented as a hypothetical structure to complete the series - the pure silicon 111 surface is not stable in the $\sqrt{3}\times\sqrt{3}$ phase, but reorders to a very large $7\times 7$ unit cell.
\end{thebibliography}
\end{document}